\documentclass[12pt]{article}

\usepackage{graphicx}
\usepackage{amssymb}
\usepackage{amsmath}
\usepackage{dsfont}
\usepackage{bm}
\usepackage{mathrsfs}

\begin{document}
\title{Quantum levitation by left-handed metamaterials}
\author{Ulf Leonhardt and Thomas G. Philbin\\
School of Physics and Astronomy, University of St Andrews,\\
North Haugh, St Andrews KY16 9SS, Scotland
}
\date{\today}
\maketitle
\begin{abstract}
Left-handed metamaterials make perfect lenses that image classical electromagnetic fields with significantly higher resolution than the diffraction limit. Here we consider the quantum physics of such devices. We show that the Casimir force of two conducting plates may turn from attraction to repulsion if a perfect lens is sandwiched between them. For optical left-handed metamaterials this repulsive force of the quantum vacuum may levitate ultra-thin mirrors.
\end{abstract}

\newpage

Left-handed metamaterials \cite{Smith}-\cite{MilonniLeft} are known to make perfect lenses \cite{Pendry} that image classical electromagnetic fields with higher resolution than the diffraction limit. Here we consider the quantum physics of such devices, in particular how they modify the zero-point energy of the electromagnetic field and the resulting mechanical force of the quantum vacuum, the Casimir force \cite{Casimir0}-\cite{Chan}. We show that the Casimir force of two conducting plates may turn from attraction to repulsion if a perfect lens is sandwiched between them. For optical left-handed metamaterials \cite{Soukoulis}-\cite{Grigorenko}, this repulsive force of the quantum vacuum may levitate ultra-thin mirrors on, literally, nothing. The usually attractive vacuum forces  \cite{Milonni}, the Casimir and the related van-der-Waals force \cite{Milonni}, are significant on the length scale of nanomachines \cite{Chan}; so ideas for manipulating vacuum forces may find applications in nanotechnology \cite{Ball}.

Repulsive Casimir forces 
\cite{Buks}-\cite{Henkel}
have been predicted to occur 
between two different extended dielectric plates,
in the extreme case \cite{Boyer0}
between one dielectric with infinite electric permittivity $\varepsilon$
and another one with infinite magnetic permeability $\mu$.
They have not been practical yet and
are subject to controversy
\cite{Iannuzzi,KKMRreply}.
The closely related case of repulsive van der Waals forces 
has been studied as well \cite{Buhmann}.
Here we consider a different situation 
inspired by Casimir's original idea \cite{Casimir0}:
imagine instead of two extended dielectric plates
two perfect conductors
with a metamaterial sandwiched in between
for which $\varepsilon=\mu=-1$
(Fig. 1A).
Such materials can be made of nanofabricated metal structures 
\cite{Soukoulis}-\cite{Grigorenko}.
One of the conducting plates may be very thin and movable,
which gives a key advantage in observing 
repulsive vacuum forces.

\begin{figure}[h]
\begin{center}
\includegraphics[width=20.0pc]{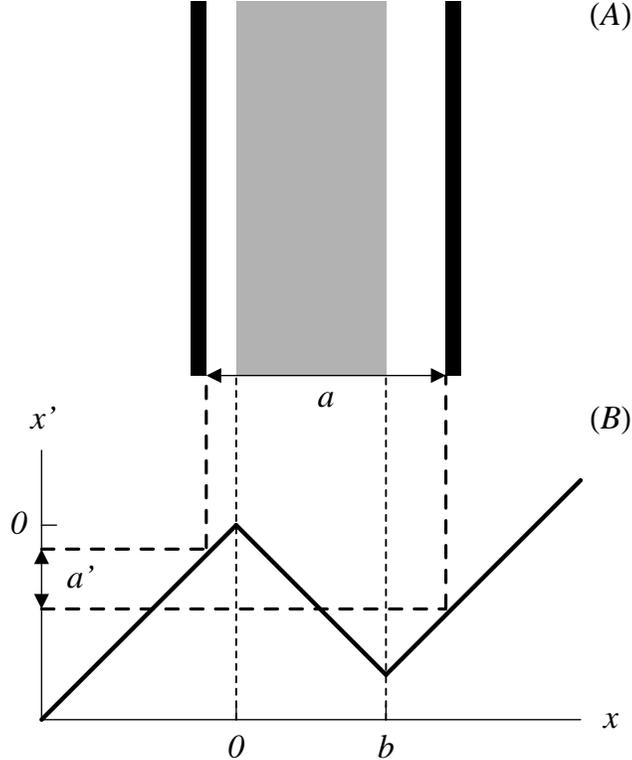}
\caption{
\small{
Casimir effect of left-handed metamaterials.
(A) illustrates a material with $\varepsilon=\mu=-1$
sandwiched between two mirrors.
(B) shows how the medium transforms 
the Casimir cavity of size $a$ in physical $x$ space
into a cavity in $x'$ space of size $a'$
according to equation (\ref{eq:left}).
The attractive Casimir force in $x'$ space
moves the mirrors further apart in $x$ space:
the Casimir effect in physical space is repulsive.
}
\label{fig:left}}
\end{center}
\end{figure}

In the following we develop a visual argument 
why the Casimir force in this set-up is repulsive.
We utilize the fact that a slab of a metamaterial 
with $\varepsilon=\mu=-1$
acts as a transformation medium \cite{LeoPhil}.
Transformation media \cite{LeoPhil,PSS}
map electromagnetic fields in physical space
to the electromagnetism of empty flat space.\footnote{
A quantum theory of light in spatial 
transformation media 
has been developed in Ref.\ \cite{LPQuantum}.}
Such media are at the heart of macroscopic
invisibility devices 
\cite{PSS}-\cite{Schurig}.
Note that electromagnetic analogues of the event horizon
\cite{LeoReview,SchUh}  
are also manifestations of transformation media \cite{LeoPhil}.

A transformation medium performs an active coordinate 
transformation:
electromagnetism in physical space,
including the effect of the medium,
is equivalent to electromagnetism in 
transformed coordinates 
where space appears to be empty.
The sole function of the device is to facilitate this
transformation.
For example, a metamaterial 
with $\varepsilon=\mu=-1$ of thickness $b$ (Fig. 1A)
transforms the
Cartesian coordinate $x$ into $x'$ as  \cite{LeoPhil} (Fig. 1B)
\begin{equation}
x' =
\left\{
\begin{array}{rcl}
x & \mbox{for} & x<0\\
-x & \mbox{for} & 0 \le x \le b\\
x-2b & \mbox{for} &  x>b
\end{array}
\right. \,.
\label{eq:left}
\end{equation}
This transformation property of the medium
visually explains \cite{LeoPhil}
that the material acts as a perfect lens 
\cite{Pendry}:
the electromagnetic field in the range  $-b<x<0$
is mapped into $x'$, but $x'$ has two more images
in physical space, one inside the device and one
in $b<x<2b$.
Since these are faithful transformations the images are perfect.
Note that the transformed coordinate system 
$\{x',y,z\}$ is left-handed \cite{LeoPhil} within the material,
which also explains why
media with  $\varepsilon=\mu=-1$
create left-handed electromagnetism 
\cite{Smith,Veselago}.
Hence they are called left-handed metamaterials 
\cite{MilonniLeft}.

Suppose that this medium is sandwiched between a Casimir 
cavity of two perfect conductors with distance $a$ (Fig. 1A).
In transformed space the plate-distance appears as  (Fig. 1B)
\begin{equation}
a' = |a-2b| \,.
\label{eq:a}
\end{equation}
Suppose that $a<2b$, {\it i.e.} 
the Casimir plates lie within the imaging range 
of the perfect lens.
In this case, 
the cavity in physical space increases when
the transformed cavity decreases.
Consequently, the attractive Casimir force
in transformed space turns into a repulsive force
in physical space.

In order to derive a quantitative result
for the Casimir force,
we note that the electromagnetic spectrum $\{\omega_\nu\}$
of the left-handed Casimir cavity 
is the spectrum of an empty cavity with distance $a'$
where $\nu$ refers to all possible modes that fit 
into the transformed cavity.
Each mode corresponds to a quantum harmonic oscillator
with ground-state energy $\hbar\omega_\nu/2$.
The total zero-point energy $U(a')$, the sum of all 
$\hbar\omega_\nu/2$, is infinite,
but differences between $U$ per area, ${\cal U}$,
for different plate distances are finite 
\cite{Milonni}-\cite{Bordag}.
The gradient of the zero-point energy causes a force
that becomes observable when one plate is movable, 
the Casimir force \cite{Casimir0}.
For empty cavities, this force is well-known 
and has been experimentally observed \cite{Chan}.
In our case, the cavity has an effective 
length of $a'$, but a mechanical length of $a$.
We use the standard result for Casimir cavities
\cite{Casimir0}
and obtain the force per area
\begin{equation}
f  = -\frac{\partial \cal U}{\partial a'}\frac{\partial a'}{\partial a} 
= \frac{\hbar c \pi^2}{240 a'^4} \,.
\label{eq:casimir}
\end{equation}

In our argument we made the implicit assumption
that the metamaterial performs 
the left-handed transformation (\ref{eq:left})
on electromagnetic modes
over a sufficient frequency range,
whereas
the negatively-refracting metamaterials of present technology
have been highly dispersive 
\cite{Soukoulis}-\cite{Grigorenko}.
In the Appendix we consider the Lifshitz theory 
\cite{LL9}-\cite{Raabe}
of the Casimir effect in dispersive materials.
We find that the transformation medium is only required
to act as a perfect lens for purely imaginary frequencies
that correspond to wavelengths comparable to or larger than  $2a'$.
However, such imaginary perfect lenses
are only possible in media with gain \cite{LL5},
{\it i.e.} in active metamaterials.
We give an example that agrees very well 
with our simple result (\ref{eq:casimir}).
We do not expect that the force (\ref{eq:casimir}) diverges
for $a' \rightarrow 0$,
when the perfect lens images 
the two plates into each other,
because no medium can sustain 
$\varepsilon \sim \mu \sim -1$
for arbitrarily short or arbitrarily long wavelengths \cite{LL8}.
We also show in the Appendix
that passive metamaterials with negative $\mu$,
but positive $\varepsilon$, 
exhibit a repulsive Casimir force,
that quantitatively differs from formula (\ref{eq:casimir}),
but not qualitatively. 
Such metamaterials can be made with present technology.

We may speculate how strong the repelling force of the
quantum vacuum may become.
We give a rough estimate that indicates
that one might levitate one of the mirrors of the 
Casimir cavity (Fig. 2).
Suppose this mirror is a $0.5\mu\mathrm{m}$ thin
pure aluminum foil in vacuum
(much thicker than 
the optical skin depth of aluminum such 
that it acts as a mirror.)
On this foil, with density
$2700\,\mathrm{kg}/\mathrm{m}^3$,
the Earth's gravity would exert a force
per area of about $0.013\mathrm{N}/\mathrm{m}^2$. 
Assuming the left-handed medium operates 
for electromagnetic modes of around $1\mu\mathrm{m}$
wavelength  \cite{Zhang}-\cite{Grigorenko},
the effective plate-distance $a'$ of the
transformed Casimir cavity can be as small as
$0.5\mu\mathrm{m}$.
In this case, the repulsive Casimir force (\ref{eq:casimir})
balances the weight of the aluminum foil:
the foil would levitate, 
carried by zero-point fluctuations.

{\it Note added.} 
Capasso informed us that a repulsive Casimir force exists
between three dielectrics 
(for example Silicon, Ethanol and Gold)
with $\mu=1$
if $\varepsilon_1-\varepsilon_3$
and $\varepsilon_2-\varepsilon_3$
have different signs
in the relevant range
of purely imaginary frequencies
\cite{BLP}.

\begin{figure}[h]
\begin{center}
\includegraphics[width=20.0pc]{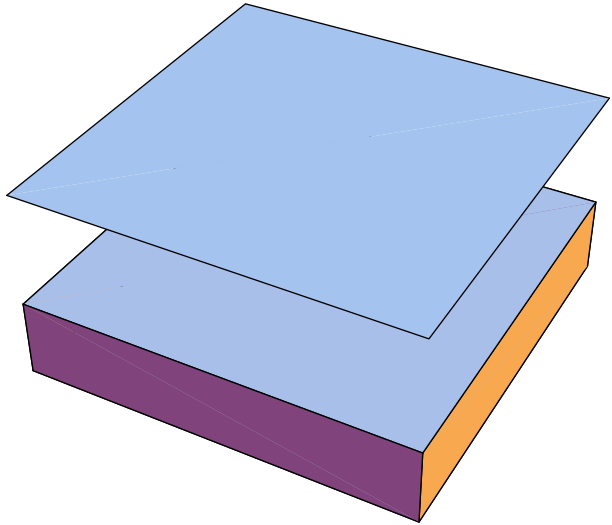}
\caption{
\small{
Levitating mirror. The repulsive Casimir force
of a left-handed material may balance the weight
of one of the mirrors, letting it levitate
on zero-point fluctuations.
}
\label{fig:levitation}}
\end{center}
\end{figure}

\section*{Acknowledgments}

We thank 
J. W. Allen, F. Capasso, M. Killi, J. B. Pendry, S. Scheel and T. Tyc 
for their comments
and the Leverhulme Trust for financial support. 

\newpage

\renewcommand{\theequation}{A\arabic{equation}}
\setcounter{equation}{0}

\section{Appendix}

In this appendix, 
we calculate the Casimir force for a dispersive medium in our set-up. 
For this, we use the generalization \cite{Raabe}
of Lifshitz' classic theory of the Casimir force in media 
\cite{LL9}-\cite{Rodriguez}. 
First, we express the electromagnetic force density ${\mathbf f}$ 
of the quantum vacuum
in terms of the quantum expectation value $\sigma$ of 
Maxwell's stress tensor \cite{Jackson},
\begin{equation}
{\mathbf f} = \nabla \cdot \sigma \, .
\end{equation}
The vacuum stress turns out to be infinite, but not its divergence, if we adopt the following regularization procedure:
consider instead of the expectation value of the stress tensor the more general correlation function
\begin{equation}
\sigma \left( {\mathbf r}, {\mathbf r'} \right) = \tau \left( {\mathbf r}, {\mathbf r'} \right) - \frac{1}{2} \, {\rm Tr} \, \tau \left( {\mathbf r}, {\mathbf r'} \right) \mathds{1}
\label{sigma}
\end{equation}
where $\mathds{1}$ denotes the three--dimensional unity matrix, $\otimes$ describes the tensor product of two vectors (a matrix ${\mathbf F} \otimes {\mathbf G}$ with components $F_l \, G_m$) and
${\rm Tr}$ denotes the trace of three-dimensional matrices. 
The matrix $\tau$ is defined as the expectation value
\begin{equation}
\tau \left( {\mathbf r}, {\mathbf r'} \right) = \varepsilon_0 \langle 0 | \hat{\mathbf E} ({\mathbf r}) \otimes \hat{\mathbf E} ({\mathbf r'}) + 
c^2 \hat{\mathbf B} ({\mathbf r}) \otimes \hat{\mathbf B} ({\mathbf r'}) | 0 \rangle
\label{tau}
\end{equation}
where $\varepsilon_0$ denotes the permittivity of the vacuum.
As we will see, $\sigma ( {\mathbf r}, {\mathbf r'} )$ is finite for ${\mathbf r} \neq {\mathbf r'}$ and approaches infinity for ${\mathbf r} \to {\mathbf r'}$, but the infinite contribution $\sigma_{\infty} ( {\mathbf r}, {\mathbf r'} )$ to $\sigma ( {\mathbf r}, {\mathbf r'} )$ does not exert any electromagnetic force, because $\nabla \cdot \sigma_{\infty} ( {\mathbf r}, {\mathbf r'} )$ turns out to vanish.
Therefore, the part of the vacuum stress that really appears as a force is finite and given by
\begin{equation}
\sigma \left( {\mathbf r} \right) = \lim_{{\mathbf r}\to{\mathbf r'}}
\Big(  \sigma ( {\mathbf r}, {\mathbf r'} ) - \sigma_{\infty} ( {\mathbf r}, {\mathbf r'} ) \Big) \, .
\end{equation}
Second,
we express the correlation functions $\langle 0 | \hat{\mathbf E} ({\mathbf r}) \otimes \hat{\mathbf E} ({\mathbf r'}) | 0 \rangle $ and $\langle 0 | 
\hat{\mathbf B} ({\mathbf r}) \otimes \hat{\mathbf B} ({\mathbf r'}) | 0 \rangle $ in terms of the classical electromagnetic Green's function
$G ({\mathbf r}, {\mathbf r'}, \omega) $.
The Green's function is proportional to the electric field ${\mathbf E} ({\mathbf r})$ of a single external dipole placed at position ${\mathbf r'}$ where it oscillates with frequency $\omega$. 
The dipole generates electromagnetic radiation that probes the properties of the medium.
This dipolar probe may point in three possible spatial directions; 
and so the Green's function is a matrix where each column corresponds to any of the directions of the dipole.
Mathematically, the Green's function is defined as the solution of the inhomogeneous electromagnetic wave equation
\begin{equation}
\nabla \times \mu^{-1}\, \nabla \times G - \varepsilon \frac{\omega^2}{c^2} G = \mathds{1} \, \delta ( {\mathbf r} - {\mathbf r'} ) 
\label{gwave}
\end{equation}
where, in general, the Fourier-transformed electric permittivity $\varepsilon$ and magnetic permeability $\mu$ are complex functions of $\omega$.  
In the limit of large frequencies $\varepsilon$ and  $\mu$ tend to unity --- 
all media are transparent to electromagnetic waves with extremely short wavelengths.
Both $\varepsilon$ and $\mu$ satisfy the crossing relations
\begin{equation}
\varepsilon(-\omega^*) = \varepsilon^*( \omega)   \, , \quad
\mu(-\omega^*) = \mu^*( \omega) 
\label{cross}
\end{equation}
and, due to causality \cite{LL8},
the dielectric functions are analytic on the upper half plane.

According to the quantum theory of light in 
dispersive and dissipative media 
\cite{KSW}-\cite{BW},
we can express
the correlations functions (\ref{tau}) in the vacuum stress (\ref{sigma})
as \cite{Raabe}
\begin{eqnarray}
\langle 0 | \hat{\mathbf E} ({\mathbf r}) \otimes \hat{\mathbf E} ({\mathbf r'}) | 0 \rangle &=&  - \frac{\hbar}{\varepsilon_0 c^2 \pi} \,  
\int_0^{\infty} \xi^2 \, G ({\mathbf r}, {\mathbf r'}, {\rm i} \xi) \,  \mathrm{d} \xi \,, 
\nonumber\\
\langle 0 | \hat{\mathbf B} ({\mathbf r}) \otimes \hat{\mathbf B} ({\mathbf r'}) | 0 \rangle &=&  \frac{\hbar}{\varepsilon_0 c^2 \pi} \,  
\int_0^{\infty} \nabla \times G ({\mathbf r}, {\mathbf r'}, {\rm i} \xi) 
\times \stackrel{\longleftarrow}{\nabla'}  \mathrm{d} \xi 
\label{eebb}
\end{eqnarray}
where $\stackrel{\longleftarrow}{\nabla'}$ indicates that differentiations are performed  from the right.
The Green's function is evaluated for purely imaginary frequencies $\mathrm{i}\xi$,
because here the integrals (\ref{eebb}) converge nicely, in general.
Note that for purely imaginary frequencies the dielectric functions
are real, because of the crossing property (\ref{cross}),
and so is the Green's function.

Consider a uniform medium where $\varepsilon$ and $\mu$ are constant in space.
In this case, we represent the electromagnetic Green's function $G ({\mathbf r}, {\mathbf r'}, {\rm i} \xi)$ in terms of the scalar Green's function $g$ that obeys the inhomogeneous wave equation
\begin{equation}
\xi^2 c^{-2} \left( \nabla^2 - \xi^2 c^{-2} \right)g = \delta ({\mathbf r} - {\mathbf r'}) \,.
\label{gdef}
\end{equation}
We use the identity 
\begin{equation}
\left( \nabla \times \nabla \times + \, \kappa^2 \mathds{1} \right)
\left( \nabla \otimes \nabla - \, \kappa^2 \mathds{1} \right) =
\kappa^2 \mathds{1} \left( \nabla^2 - \kappa^2 \, \right)
\end{equation}
for $\kappa^2 = \varepsilon \mu \xi^2 c^{-2}$, 
and see that the solution of the electromagnetic wave equation (\ref{gwave}) is 
\begin{equation}
G = \sqrt{\frac{\mu}{\varepsilon}} \left( \nabla \otimes \nabla - \, \varepsilon \mu \xi^2 c^{-2} \mathds{1} \right)
g \left( \sqrt{\varepsilon \mu} ({\mathbf r} - {\mathbf r'}) \right) \, .
\label{g0}
\end{equation}
Uniform media do not feel the force of the quantum vacuum
--- otherwise they would disintegrate; but
the Green's function and hence the vacuum stress approaches infinity for ${\mathbf r} \rightarrow {\mathbf r'}$.
The infinity of the Green's function for uniform media characterizes the infinity in the non-uniform case, because, around the singularity of the delta function in the wave equation (\ref{gwave}) we can assume the medium to be uniform; 
the dominant, diverging contribution to $G$ is given by the uniform Green's function (\ref{g0}) with $\varepsilon = \varepsilon ({\mathbf r'})$ and $\mu = \mu ({\mathbf r'})$.
On the other hand, we know that this contribution does not generate a force.
In this way, we arrive at a simple recipe for removing the most severely infinite but physically insignificant contribution $\sigma_{\infty} ({\mathbf r}, {\mathbf r'})$ from the correlation function (\ref{sigma}): 
$\sigma_{\infty} ({\mathbf r}, {\mathbf r'})$ is constructed from the uniform Green's function (\ref{g0}) taking the local values of $\varepsilon$ and $\mu$ at ${\mathbf{r}}'$.
Equivalently, we can remove the uniform Green's functions from the integrals (\ref{eebb}) without changing the vacuum force. 

Consider the Casimir set-up illustrated in Fig.\ \ref{fig:left} 
with a left-handed metamaterial  sandwiched between the two mirrors.
In order to describe the influence of dispersion in the simplest possible way,
we use a simple toy model: we assume that the medium acts as a transformation medium for all frequencies, 
\begin{equation}
x' =
\left\{
\begin{array}{rcl}
x & \mbox{for} & x<0\\
x/\varepsilon(\omega) & \mbox{for} & 0 \le x \le b\\
x-b+b/\varepsilon(\omega) & \mbox{for} &  x>b
\end{array}
\right. \,,
\label{trans}
\end{equation}
but the transformation depends on frequency.
Physically, this medium corresponds to the 
impedance-matched permittivity and permeability tensors
\cite{LeoPhil}
\begin{equation}
\varepsilon^j_k = \mu^j_k = 
\mathrm{diag}
\left(\frac{\mathrm{d}x}{\mathrm{d}x'},
\frac{\mathrm{d}x'}{\mathrm{d}x},
\frac{\mathrm{d}x'}{\mathrm{d}x}\right)
= 
\left\{
\begin{array}{rl}
\mathds{1} & \mbox{outside}\\
\mathrm{diag}\left(\varepsilon,\varepsilon^{-1},\varepsilon^{-1}\right)& \mbox{inside}
\end{array}
\right.
\,.
\label{lens}
\end{equation}
 Although the medium (\ref{lens}) performs the coordinate transformation (\ref{trans}), the Green's function $G$ is not the transformed Green's 
function of the empty Casimir cavity, 
because the source on the right-hand side of
the inhomogeneous electromagnetic wave equation (\ref{gwave}),
the current of the dipolar probe,
is not transformed according to the rules \cite{LeoPhil}. For example the Green's function (\ref{gwave}) for the infinitely extended anisotropic transformation medium (\ref{lens}) (without the cavity plates) is
\begin{eqnarray}
G_{xx}&=&\pm\left(\partial_x\otimes\partial_x-\xi^2c^{-2}\varepsilon^{-2}\right)g\,, \quad
G_{xA}=G_{Ax}=\pm\partial_x\otimes\partial_Ag\,, \nonumber \\
G_{AB}&=&\pm\left(\partial_A\otimes\partial_B-\xi^2c^{-2}\delta_{AB}\right)g\,, \quad
A,B=\{y,z\}\,, \label{Gtrans}
\end{eqnarray}
where $g$ is written in transformed coordinates $x\rightarrow x/\varepsilon$, and the $\pm$ refers to the sign of $\varepsilon$. The Green's function (\ref{Gtrans}) is not the transformed uniform Green's function (\ref{g0}). Nevertheless, after solving the wave equation (\ref{gwave}) with the boundary conditions at the cavity plates, subtracting the vacuum Green's function and calculating the correlation functions (\ref{eebb}) we obtain the surprisingly simple result
\begin{equation}
\sigma_{xx}
= \frac{\hbar}{\pi^2} \int_{0}^{\infty} \int_{0}^{\infty} 
\frac{\displaystyle w u }
{\displaystyle {\rm e}^{2a'w}-1} \, {\rm d}u \,{\rm d}\xi
\,,\quad
w^2 = u^2 + \xi^2c^{-2} \,,
\label{lifshitz}
\end{equation}
Lifshitz' formula \cite{LL9}
in terms of the transformed cavity distance
\begin{equation}
a' = a-b + \frac{b}{\varepsilon({\rm i}\xi)} \,.
\label{size}
\end{equation}
The medium thus changes the effective
cavity size in the Casimir stress according to the 
transformations (\ref{trans});
but, of course, the cavity size $a'$ depends on frequency. 
Note that the off-diagonal components of the stress tensor
vanish due to the symmetry of our set-up. 
Inside the Casimir cavity, the vacuum stress is constant, but not isotropic.
Outside of the cavity, the Green's function $G$ vanishes, and so does the stress tensor.
Consequently, $\sigma$ jumps from zero to diag$(\sigma_{xx}, \sigma_{yy}, \sigma_{zz})$ at the inner surface of the Casimir plates.
The force density $\nabla \cdot \sigma$ gives a delta function at the surface and points in positive $x$ direction at the left plate and in negative $x$ direction at the right plate, if $a'$ is positive.
In this case,
the vacuum stress causes an attractive force towards the interior.

In the case of negative $\varepsilon({\rm i}\xi)$
the effective cavity size (\ref{size}) is negative
and so the Lifshitz integral (\ref{lifshitz}) diverges,
despite our regularization procedure that has removed 
the principal singularity of the vacuum stress; 
but this additional divergence of (\ref{lifshitz}) does 
not contribute to the Casimir force either, because
\begin{equation}
\frac{\displaystyle wu}
{\displaystyle {\rm e}^{2a'w}-1} 
= -\frac{\displaystyle wu}{\displaystyle {\rm e}^{-2a'w}-1} - wu \,;
\end{equation}
the integral of the first term converges and
the diverging integral of $-wu$ does not depend on the cavity at all.
Furthermore, the Casimir force changes sign, 
from attraction to repulsion, in agreement with
our simple argument in the main part of this paper.

\begin{figure}[h]
\begin{center}
\includegraphics[width=20.0pc]{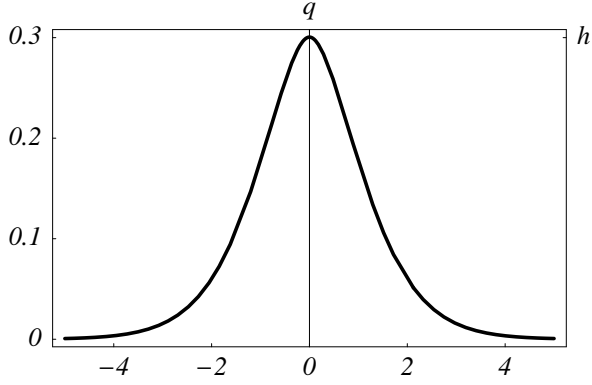}
\caption{
\small{
Lifshitz kernel.
The figure shows the kernel (\ref{kernel}) used in the 
Lifshitz theory of the Casimir effect in dispersive materials.
}
\label{fig:poly}}
\end{center}
\end{figure}

Adopting this additional regularization procedure in spectral regions
where $a'$ is negative, 
we express the Casimir force as
\begin{equation}
f = -\frac{\hbar}{\pi^2} \int_{0}^{\infty}  
\frac{h(q)}{a'^3} \, {\rm d}\xi
\,,\quad
q= |a'| \xi c^{-1} \,.
\label{cforce}
\end{equation}
To solve
the remaining integral for $h(q)$,
we use Eq.\ 2.3.14.5 of Ref.\ \cite{Prudnikov}, Vol.\ I, and obtain
\begin{equation}
h = \frac{1}{4}\mathrm{Li}_3({\rm e}^{-2q}) + \frac{|q|}{2}\,\mathrm{Li}_2({\rm e}^{-2q}) - \frac{q^2}{2} \ln (1- {\rm e}^{-2q})
\label{kernel}
\end{equation}
in terms of the polylogarithms 
$\mathrm{Li}_n(z)= \sum_{k=1}^\infty {z^k}/{k^n}$.
Figure \ref{fig:poly} shows that the kernel 
$h(q)$ is peaked around zero with a width of roughly $\pi$.
Consequently, if $a'$ is negative, 
but does not vary much for purely imaginary frequencies in the spectral region until $\pi c/|a'|$, 
we obtain the repulse Casimir force (\ref{eq:casimir}),
the main quantitative result of this paper.

\begin{figure}[b]
\begin{center}
\includegraphics[width=20.0pc]{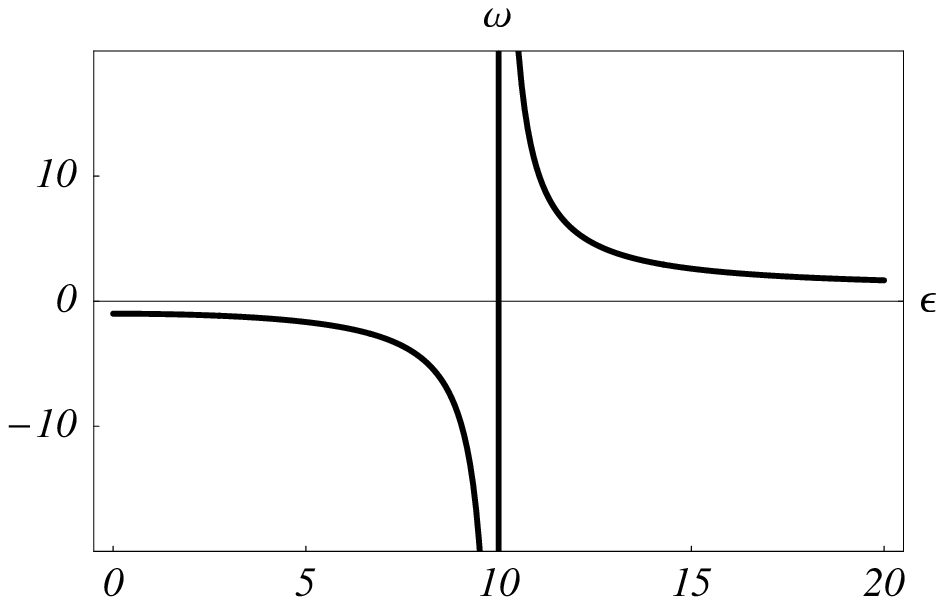}
\includegraphics[width=20.0pc]{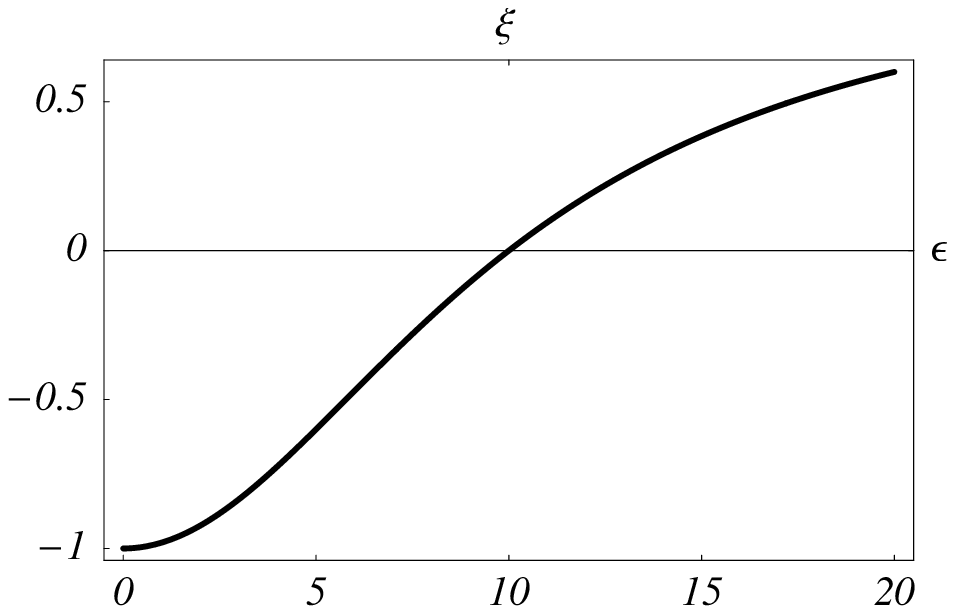}
\caption{
\small{
Gain line. The figures show the electric permittivity $\varepsilon$
of an active medium with a single spectral line of gain.
We use the Drude formula (\ref{drude}) in dimensionless units
with the resonance frequency $\omega_0=10$.
The upper plot shows $\varepsilon$ for real frequencies,
whereas the lower plot shows $\varepsilon$ 
for purely imaginary frequencies $\mathrm{i}\xi$.
On the imaginary axis, $\varepsilon$ is close to $-1$
in a sufficiently long frequency interval.
As long as this interval lies within $(0,\pi c/|a'|)$
the Casimir force agrees very well with the
simple expression (\ref{eq:casimir}), 
see also Fig.\ \ref{fig:force}.
The behavior of  $\varepsilon$ for large imaginary frequencies 
is significantly less relevant to the Casimir force. 
}
\label{fig:drude}}
\end{center}
\end{figure}

\begin{figure}[h]
\begin{center}
\includegraphics[width=20.0pc]{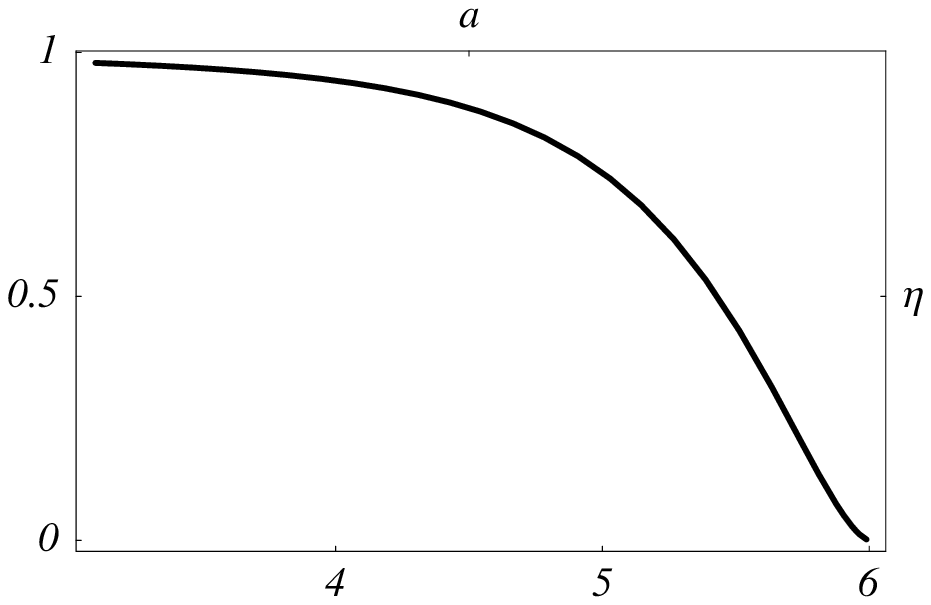}
\caption{
\small{
Comparison. 
In order to compare the simple formula (\ref{eq:casimir}) 
for the Casimir force with the result (\ref{cforce})
of the more sophisticated Lifshitz theory,
we plot the ratio $\eta$ between the 
Lifshitz force (\ref{cforce}) and the Casimir formula
(\ref{eq:casimir}) as a function of the 
cavity size $a$ (in real space) for $b=3$
in dimensionless units.
}
\label{fig:force}}
\end{center}
\end{figure}

However, negative $\varepsilon({\rm i}\xi)$ for positively imaginary frequencies are impossible in absorptive dielectrics \cite{LL5}; 
they can only occur in media with gain. 
In absorptive media the imaginary part of $\varepsilon$ 
is positive for real frequencies $\omega$.
In this case, causality --- the analyticity of $\varepsilon(\omega)$ on the upper half plane --- implies that $\varepsilon({\rm i}\xi)$ is positive, 
see Ref.\  \cite{LL5}, \S123.
On the other hand, consider a medium with a single spectral line of gain described by the simple Drude model \cite{Jackson}
\begin{equation}
\varepsilon(\omega) = 1 - \frac{2\omega_0^2}{\omega_0^2-\omega^2 -\mathrm{i}\gamma\omega}
\label{drude}
\end{equation}
in the limit of vanishing, but positive gain $\gamma$.
Figure \ref{fig:drude} illustrates the resonance in $\varepsilon(\omega)$ and the behavior of $\varepsilon({\rm i}\xi)$ for positively imaginary frequencies
where $\varepsilon$ remains $-1$ over a sufficiently large range of the spectrum.
Figure \ref{fig:force} illustrates the remarkable accuracy of the simple prediction (\ref{eq:casimir}) for the Casimir force in comparison with the Lifshitz theory for dispersive dielectrics with gain. 

One might object that the permittivity tensor (\ref{lens}) for a transformation medium 
has singularities on the upper half plane; and hence the medium is not causal. 
However,
all that really enters our simple model is the assumption that 
the medium performs the transformation (\ref{trans}) over the relevant range of purely imaginary frequencies; 
the medium may deviate from the transformation rule  (\ref{lens}) in the vicinity of a zero in $\varepsilon({\rm i}\xi)$,
which results in a more complicated expression for the vacuum stress, 
but produces, in very good approximation, the same result.

\begin{figure}[h]
\begin{center}
\includegraphics[width=20.0pc]{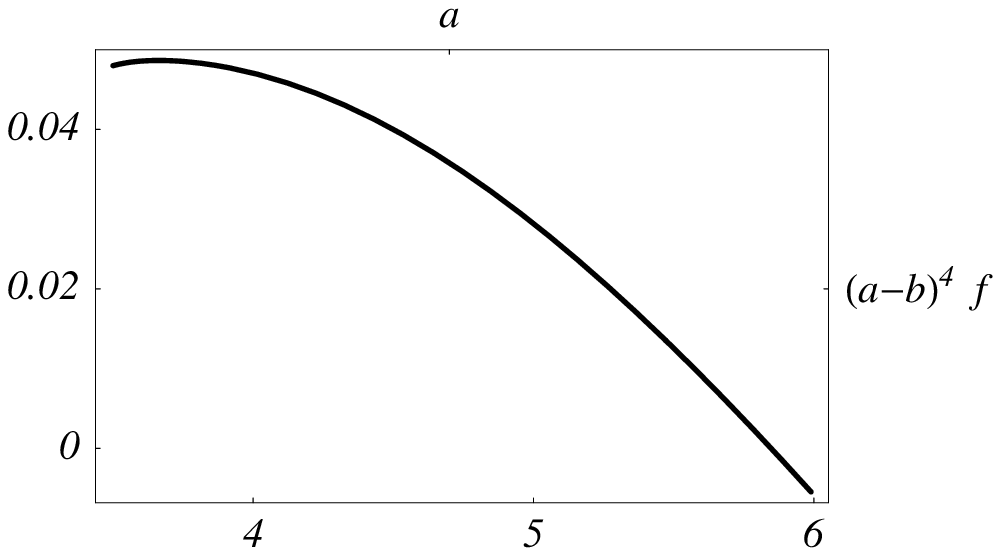}
\caption{
\small{
Magnetic materials. Casimir force for the arrangement of Fig.~\ref{fig:levitation} with $\varepsilon$ and $\mu$ given by (\ref{magnetic}), with $\Omega=5$ and $\omega_0=10$ in dimensionless units. The force $f$ on the upper plate is calculated from the stress (\ref{magstress})  and is re-scaled by dividing by the Casimir factor $\hbar c\pi^2/240$. The plot shows $f$ multiplied by the fourth power of the distance $a-b$ between the plate and the material versus the cavity size $a$ for $b=3$.
}
\label{fig:magnetic}}
\end{center}
\end{figure}

Moreover, we can relax the assumption that the medium is impedance-matched and still obtain a repulsive Casimir force, even without gain, as long as  $\mu({\rm i}\xi)$ is greater than $\varepsilon({\rm i}\xi)$ for purely imaginary frequencies over a sufficiently large spectral range. This is achieved, for example, by setting
\begin{equation} 
\varepsilon=1\,, \quad 
\mu(\omega)=1 + \frac{\Omega^2}{\omega_0^2-\omega^2} \,,
\label{magnetic}
\end{equation}
so that there is just a magnetic response in the material, the permeability being given by a Drude formula. If we consider the geometrical arrangement of Fig.~\ref{fig:levitation}, where the material is in contact with the lower mirror, then the vertical vacuum stress at the upper mirror turns out to be
\begin{eqnarray}
\sigma_{xx}&=&\frac{\hbar}{\pi^2} \int_{0}^{\infty} \int_{0}^{\infty} 
uw \varrho\,
{\rm d}u \,{\rm d}\xi
\,,
\nonumber\\
\varrho & = &\frac{ sw\left[ \varepsilon  + \mu+
       \left( K^2 + L^2 \right)\left( \varepsilon  - \mu  \right)    \right]  + 
      2\,e^{-2\left(a- b  \right) w}
       \left(  Lw\varepsilon  - Ks \right) 
       \left(  Kw\mu  - Ls    \right)  }{8
    \left( KNs + LMw\varepsilon  \right) \left( LMs + KNw\mu  \right) }
\,,
\nonumber\\  
K&=&\sinh(bs), \quad L=\cosh(bs), \quad M=\sinh\left[(a-b)w\right], \quad N=\cosh\left[(a-b)w\right]
\,,
\nonumber\\      
w^2 &=& u^2 + \xi^2c^{-2} \,, \quad s^2 = u^2 + \varepsilon\mu\xi^2c^{-2} 
\,. \label{magstress}
\end{eqnarray}
Figure\ \ref{fig:magnetic} shows that the Casimir force is repulsive when the distance $a-b$
between the material and the upper mirror is of the order $\pi c/\omega_0$, where $\omega_0$ is the resonance frequency in the Drude formula (\ref{magnetic}).

\begin{thebibliography}{99}
\bibitem{Smith}
D. R. Smith, J. B. Pendry, and M. C. K. Wiltshire,
Science {\bf 305}, 788 (2004).
\bibitem{Veselago}
V. Veselago {\it et al.},
J. Comp. Theor. Nanoscience {\bf 3}, 189 (2006).
\bibitem{MilonniLeft}
P. W. Milonni,
{\it Fast Light, Slow Light and Left-Handed Light}
(Taylor and Francis, New York, 2005).
\bibitem{Pendry}
J. B. Pendry, 
Phys. Rev. Lett. {\bf 85}, 3966 (2000).
\bibitem{Casimir0}
H. B. G. Casimir,
Proc. Kon. Ned. Akad. Wetenschap. {\bf 51}, 793 (1948).
\bibitem{Milonni}
P. W. Milonni,
{\it The Quantum Vacuum}
(Academic, London, 1994).
\bibitem{Lamoreaux2}
S. K. Lamoreaux,
Rep. Prog. Phys. {\bf 68}, 201 (2005).
\bibitem{Bordag}
M. Bordag, U. Mohideen, and V. M. Mostepanenko,
Phys. Rep, {\bf 353}, 1 (2001).
\bibitem{Chan}
H. B. Chan {\it et al.},
Science {\bf 291}, 1941 (2001).
\bibitem{Soukoulis}
C. M. Soukoulis, S. Linden, and M. Wegener,
Science {\bf 315}, 47 (2007).
\bibitem{Zhang}
S. Zhang {\it et al.},
Phys. Rev. Lett. {\bf 95}, 137404 (2005).
\bibitem{Dolling}
G. Dolling {\it et al.}, 
Opt. Lett. {\bf 30}, 3198 (2005).
\bibitem{Shalaev}
V. M. Shalaev, 
Opt. Lett. {\bf 30}, 3356 (2005).
\bibitem{Grigorenko}
A. N. Grigorenko {\it et al.},
Nature {\bf 438}, 335 (2005).
\bibitem{Ball}
Ph. Ball, 
Nature {\bf 447}, 772 (2007). 
\bibitem{Buks}
E. Buks and M. L. Roukes,
Nature {\bf 419}, 119 (2002).
\bibitem{Boyer0}
T. H. Boyer, 
Phys. Rev. A {\bf 9}, 2078 (1974).
\bibitem{KKMR}
O. Kenneth {\it et al.},
Phys. Rev. Lett. {\bf 89}, 033001 (2002).
\bibitem{Iannuzzi}
D. Iannuzzi and F. Capasso,
Phys. Rev. Lett.  {\bf 91}, 029101 (2003).
\bibitem{KKMRreply}
O. Kenneth {\it et al.},
Phys. Rev. Lett. {\bf 91}, 029101 (2003).
\bibitem{Boyer1}
T. H. Boyer, 
Am. J. Phys. {\bf 71}, 990 (2003).
\bibitem{Henkel}
C. Henkel and K. Joulain,
Europhys. Lett. {\bf 72}, 929 (2005). 
\bibitem{Buhmann}
S. Y. Buhmann, D.-G. Welsch, and T. Kampf,
Phys. Rev. A {\bf 72}, 032112 (2005).
\bibitem{LeoPhil}
U. Leonhardt and T. G. Philbin,
New J. Phys. {\bf 8}, 247 (2006).
\bibitem{PSS}
J. B. Pendry, D. Schurig, and D. R. Smith,
Science {\bf 312}, 1780 (2006).
\bibitem{LeoConform}
U. Leonhardt,
Science {\bf 312}, 1777 (2006).
\bibitem{Hendi}
A. Hendi, J. Henn, and U. Leonhardt,
Phys. Rev. Lett. {\bf 97}, 073902 (2006).
\bibitem{Schurig}
D. Schurig {\it et al.},
Science {\bf 314}, 977 (2006).
\bibitem{LPQuantum}
U. Leonhardt and T. G. Philbin,
J. Opt. A (in press).
\bibitem{LeoReview}
U. Leonhardt,  
Rep. Prog. Phys. {\bf 66}, 1207 (2003).
\bibitem{SchUh}
R. Sch\"utzhold and W. G. Unruh, 
Phys. Rev. Lett. {\bf 95}, 031301 (2005).
\bibitem{LL9}
L. D. Landau and E. M. Lifshitz,
{\it Statistical Physics. Part 2, Theory of the Condensed State}
(Pergamon, Oxford, 1980).
\bibitem{Raabe}
C. Raabe and D.-G. Welsch, 
Phys. Rev. A {\bf 71}, 013814 (2005).
\bibitem{LL5}
L. D. Landau and E. M. Lifshitz,
{\it Statistical Physics. Part 1}
(Pergamon, Oxford, 1980).
\bibitem{BLP}
I. E. Dzyaloshinskii, E. M. Lifshitz, and L. P. Pitaevskii,
Advanc. Phys. {\bf 10}, 165 (1961).
\bibitem{LL8}
L. D. Landau and E. M. Lifshitz, 
{\it Electrodynamics of Continuous Media}
(Butterworth-Heinemann, Oxford, 1993).
\bibitem{Jackson}
J. D. Jackson, 
{\it Classical Electrodynamics} (Wiley, New York, 1998).
\bibitem{Rodriguez}
A. Rodriguez {\it et al.},
arXiv:0704.1890.
\bibitem{KSW}
L. Kn\"oll, S. Scheel, and D.-G.Welsch,
{\it QED in dispersing and absorbing media}, 
in {\it Coherence and Statistics of Photons and Atoms} 
ed. by J. Perina (Wiley, New York, 2001).
\bibitem{Luks}
A. Luks and V. Perinova,
Prog. Opt. {\bf 43}, 295 (2002).
\bibitem{BW}
S. Y. Buhmann and D.-G. Welsch,
Prog. Quant. Electron. {\bf 31}, 51 (2007).
\bibitem{Prudnikov}
A. P. Prudnikov, Yu. A. Brychkov, and O. I. Marichev, 
{\it Integrals and Series}
(Gordon and Breach, New York, 1992).

\end{thebibliography}
\end{document}